\documentclass[acmtog,noacm]{acmart}

\usepackage{booktabs} %
\usepackage{amsmath,amsfonts}
\usepackage{mathtools}
\usepackage{outlines}
\usepackage{mathrsfs} %
\usepackage{siunitx}
\usepackage{ulem} %
\usepackage{graphicx} %
\usepackage{wrapfig}  %
\usepackage{lettrine}
\usepackage{float}
\usepackage{hyperref}
\usepackage{tikz,nicematrix,adjustbox}

\newcommand{\bm}[1]{{\boldsymbol{#1}}} %
\newcommand{\trsp}{{\!\scriptscriptstyle\top}}

\newcommand{\reals}{{\rm I\!R}} %
\newcommand{\cp}{{\mathbf{c}}}
\newcommand{\deriv}[1]{{\frac{\mathrm{d}}{\mathrm{d}#1}}}
\newcommand{\df}[1]{{\mathrm{d}#1}}
\newcommand{\img}{{\mathbb{I}}}

\newcommand{\refig}[1]{{Fig.~\ref{#1}}} %
\newcommand{\refsect}[1]{{Section \ref{#1}}} %

\newcommand{\mydot}{\bullet} %

\usepackage[ruled]{algorithm2e} %

\SetAlFnt{\small}
\SetAlCapFnt{\small}
\SetAlCapNameFnt{\small}
\SetAlCapHSkip{0pt}

\newcommand{\stkout}[1]{\ifmmode\text{\sout{\ensuremath{#1}}}\else\sout{#1}\fi}

\definecolor{MyGreen}{rgb}{0,0.5.2}
\definecolor{MyBlue}{rgb}{0,0.2,1.0}
\definecolor{MyRed}{rgb}{1,0.2,0.2}
\definecolor{MyPurple}{rgb}{0.5,0,1.0}
\definecolor{MyOrange}{rgb}{1.0,0.82,0}
\definecolor{MyBrown}{rgb}{0.65,0.35,0}
\definecolor{MyMagenta}{rgb}{1.,0.0,1.}
\definecolor{MyGrey}{rgb}{0.5,0.5,0.5}
\definecolor{MyLightGrey}{rgb}{0.9,0.9,0.9}

\newcommand{\dani}[1]{{\color{MyGreen}{(Daniel: #1)}}}
\newcommand{\daniel}[1]{{\color{MyGreen}{(Daniel: #1)}}}
\newcommand{\od}[1]{{\color{MyBlue}{(Oliver: #1)}}}
\newcommand{\michael}[1]{{\color{MyPurple}{(Michael: #1)}}}
\newcommand{\arik}[1]{{\color{MyRed}{(Arik: #1)}}}
\newcommand{\ffl}[1]{{\color{MyBrown}{(FFL: #1)}}}
\newcommand{\sylvain}[1]{{\color{MyOrange}{(Sylvain: #1)}}}

\newcommand{\correction}[2]{{#2}}

\newcommand{\addition}[1]{{#1}}

\newcommand{\deleted}[1]{{}}

\newcommand{\revise}[1]{{\color{MyBrown}{(Revise: #1)}}}

\newcommand{\previous}[1]{{\color{MyGrey}{(Previously: #1)}}}
\newcommand{\reviewer}[1]{{\color{MyBrown}{(Reviewer: #1)}}}
\newcommand{\fixed}[1]{{\color{MyGreen}[Fixed: #1]}}

\newcommand{\todo}[1]{{\color{MyRed}(\textbf{TODO:} #1)}}

\newcommand{\revision}[1]{{\color{MyRed}#1}}
\newcommand{\revnote}[1]{{\color{MyBlue}[Revision: #1]}}
\renewcommand{\revision}[1]{{#1}}
\renewcommand{\revnote}[1]{{}}

\ifthenelse{\isundefined{\comments}}
{
\renewcommand{\dani}[1]{}
\renewcommand{\daniel}[1]{}
\renewcommand{\arik}[1]{}
\renewcommand{\sylvain}[1]{}
\renewcommand{\od}[1]{}
\renewcommand{\michael}[1]{}
\renewcommand{\ffl}[1]{}
\renewcommand{\revise}[1]{}
\renewcommand{\previous}[1]{}
\renewcommand{\reviewer}[1]{}
\renewcommand{\correction}[2]{#2}

\renewcommand{\addition}[1]{#1}
\renewcommand{\fixed}[1]{}
\renewcommand{\deleted}[1]{}
\renewcommand{\todo}[1]{}

}
{}

\AtBeginDocument{%
  }

\setcopyright{cc}
\setcctype{by}
\acmJournal{TOG}
\acmYear{2025} \acmVolume{44} \acmNumber{6} \acmArticle{225} \acmMonth{12} \acmPrice{}\acmDOI{10.1145/3763345}

\begin{document}

\title{Neural Image Abstraction Using Long Smoothing B-Splines}

\author{Daniel Berio}
\affiliation{%
  \institution{Goldsmiths, University of London}
  \city{London}
  \country{United Kingdom}}
\email{daniel.berio@gold.ac.uk}

\author{Michael Stroh}
\affiliation{%
  \institution{University of Konstanz}
  \city{Konstanz}
  \country{Germany}
}
\email{michael.stroh@uni-konstanz.de}

\author{Sylvain Calinon}
\affiliation{%
 \institution{Idiap Research Institute}
 \city{Martigny}
 \country{Switzerland}}
\email{sylvain.calinon@idiap.ch}

\author{Frederic Fol Leymarie}
\affiliation{%
  \institution{Goldsmiths, University of London}
  \city{London}
  \country{United Kingdom}}
\email{ffl@gold.ac.uk}

\author{Oliver Deussen}
\affiliation{%
  \institution{University of Konstanz}
  \city{Konstanz}
  \country{Germany}
}
\email{oliver.deussen@uni-konstanz.de}

\author{Ariel Shamir}
\affiliation{%
  \institution{Reichman University}
  \city{Herzliya}
  \country{Israel}}
\email{arik@runi.ac.il}

\renewcommand{\shortauthors}{Berio et al.} %

\begin{abstract}
    We integrate smoothing B-splines into a standard differentiable vector graphics (DiffVG) pipeline through linear mapping, and show how this can be used to generate smooth and arbitrarily long paths within image-based deep learning systems. We take advantage of derivative-based smoothing costs for parametric control of  fidelity vs. simplicity tradeoffs, while also enabling stylization control in geometric and image spaces. The proposed pipeline is compatible with recent vector graphics generation and vectorization methods.
  We demonstrate the versatility of our approach with four applications aimed at the generation of stylized vector graphics: stylized space-filling path generation, stroke-based image abstraction, closed-area image abstraction, and stylized text generation.

\end{abstract}

\begin{teaserfigure}
    \centering
    \includegraphics[width=0.99\linewidth]{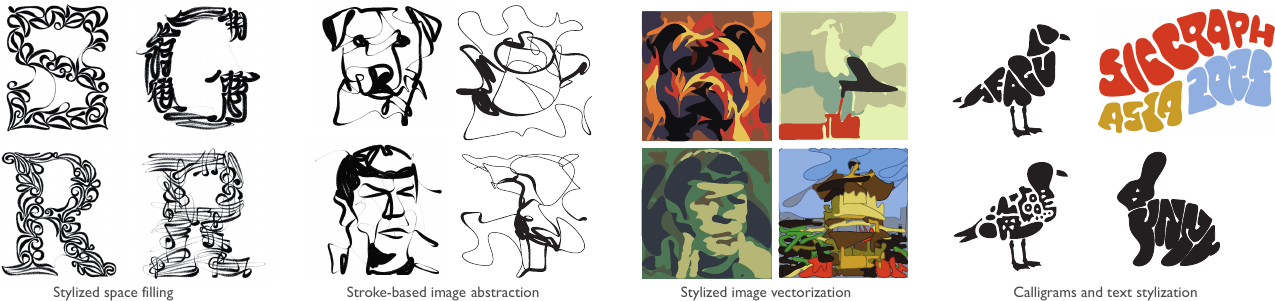}
    \caption{Our method allows to optimize long and smooth B-Spline curves using DiffVG rasterization pipelines.
      Applications range from text stylization to image abstraction and vectorization.
    }
    \label{fig:teaser}
\end{teaserfigure}




\maketitle

\section{Introduction}

\lettrine[lines=6, image=true, findent=3pt, nindent=0pt, loversize=0.06]{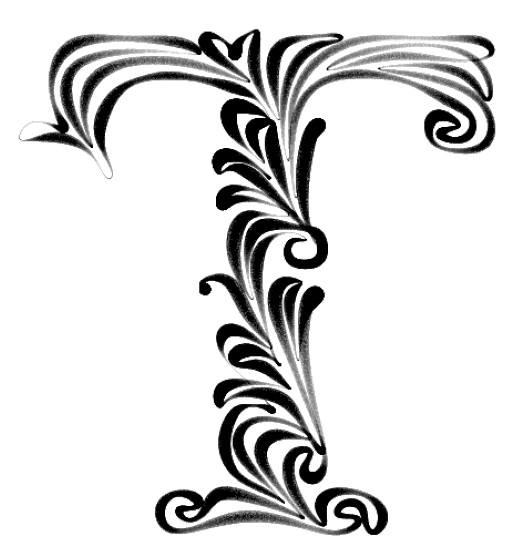}{he ability} to produce long, smooth curves is central to a variety of design and artistic tasks. These include freehand drawing, sketching, calligraphy, typography, logo design as well as image abstractions into compositions of organic, flowing, or blob\-by shapes. Our aim is to enable the generation of vector graphic outputs that allow these types of designs, while taking advantage of recent advances in gradient-based image generation, stylization, and understanding.

Developments in differentiable vector graphics (DiffVG) rasterization have enabled gradient-based optimization methods that leverage complex image-space losses to drive image generation, stylization and abstraction methods. Most existing approaches rely on the method of \citet{liDifferentiableVectorGraphics2020}, which implements differentiable rasterization for a large subset of elements of the Scalable Vector Graphics (SVG) standard, including piecewise cubic and quadratic B\'ezier curves. Most of these methods directly optimize B\'ezier curves, but even with additional smoothing penalties they do not provide guarantees of continuity across segments, which limits their ability to represent long, smooth and expressive strokes.

Our work is based on two observations. First, alternative spline para\-metrizations such as B-spline \cite{De_Boor2001-bd}  or Catmull-Rom \cite{DeRoseCatmull1988} provide inherent continuity constraints in their definition. Second, the conversion of such curves to B\'ezier curves is a linear transformation, making their integration into existing DiffVG pipelines a matter of an additional matrix multiplication. Although these curve parameterizations are well established, to the best of our knowledge, their integration into DiffVG remains largely unexplored.

In our work we focus on uniform B-splines for their simplicity, high-order continuity and analytic properties \cite{farinCurvesSurfacesCAGD2001}.
This enables a straightforward implementation of derivative-based smoothing criteria that are well known in the fairing and motor-control/robo\-tics domains, but most importantly support our goals of generating long stylized curves within a DiffVG pipeline.

We define %
B-splines with control-polygons consisting of series of ``key-points'' and convert these to piecewise cubic B\'ezier curves for rendering.
Since this transformation %
is linear and rendering is differentiable, gradients from image-space losses can be back-propa\-gated to the key-points.
We treat stroke width as a third curve dimension, where each curve control point can be assigned an independent stroke radius, enabling smooth variations similar to that seen in physical brush strokes \cite{fujiokaConstructingCharacterFont2007}. Allowing the stroke width to vanish also presents an effective way to alter the number of visible strokes required for an image abstraction. Our method operates with both open and closed curves, supporting the generation of closed and organic-looking areas.%

We present four different applications for our method: abstract space filling curves (\refsect{sec:AreaFilling}), sketch-based stylization  (\refsect{sec:ImageAbstraction}), abstract image vectorization with  color quantization (\refsect{sec:AreaAbstraction}) as well as text stylization and calligram generation with a novel legibility cost  (\refsect{sec:TextStylization}).
We provide a practical implementation of smoothing B-splines that can be directly integrated into DiffVG pipelines and demonstrate how this enables long and expressive curves while maintaining flexible geometric and stylistic control.
Working code and examples for our method are available at
\href{https://github.com/colormotor/calligraph}{\textit{\nolinkurl{github.com/colormotor/calligraph}}}.

\revnote{Removed contribuions as for R3.}

\section{Related work}
\subsection{Smooth and stylized curve generation}
\revision{Long and stylized strokes have been explored used in the literature for applications including image stylization \cite{kaplan2005tsp,Wong2011,Tong2025}, text-based stylization \cite{Maharik2011}, and fabrication \cite{Liu2017,Yang2021}. To widen and enhance such applications, our method also enables the
generation of long and smooth strokes through the use of neural-driven image-based costs. Smoothing is achieved by minimizing the squared magnitude of higher-order positional derivatives.}

\revnote{Was: Our method facilitates the implementation of smoothing criteria by minimizing the squared magnitude of higher-order positional derivatives.}

In the motor control literature, it is well established that the kinematics of hand and arm movements can be modeled by optimizing performance criteria \cite{FlashOpt98}. The so-called minimum square derivative models have been successfully applied to handwriting and curved motion by minimizing third-order derivatives (jerk) \cite{flashCoordinationArmMovements1985} and fourth-order derivatives (snap) \cite{Edelman1987}.
Similar minimum principles are widely employed for smooth motion control in drones \cite{Mellinger2011,KryDrone2019} and robots ~\cite{Toussaint17,Todorov04}, as well as in statistics for smoothing noisy data~\cite{Reinsch1967,eilersFlexibleSmoothingBsplines1996}.

Similar principles of smooth motion and continuity have also been used for curve fairing, where a ``fair'' curve is typically one that exhibits a smooth variation of the curvature \cite{farinCurvesSurfacesCAGD2001}. In this context, jerk has been adopted as an approximation for curvature variation \cite{luNoteCurvatureVariation2015,pottmannSmoothCurvesTension1990,meierInterpolatingCurvesGradual1987}, while snap serves as an approximation for transverse distributed load \cite{meierInterpolatingCurvesGradual1987}.

In an extensive body of work, \citet{egerstedtControlTheoreticSplines2009} develop ``dynamic splines'' that formulate polynomial splines through optimal control of linear systems. \citet{BerioGI2017} use similar principles for the interactive generation of stylized paths similar to the ones seen in graffiti art and calligraphy with applications similar to ours.
\citet{KanoControl2003} study the relations between dynamic splines and B-splines and in a collection of work, %
they develop an optimal formulation of B-splines \cite{kanoOptimalCurveFitting2005} applied to generate motion paths and curves similar to those found in Japanese calligraphy \cite{matsukidaModelingReshapingHandwritten2013,fujiokaConstructingReconstructingCharacters2006}. Our approach is strongly inspired by the B-spline construction initially proposed by \citet{kanoOptimalCurveFitting2005}, but we extend their formulation to support DiffVG and demonstrate its flexibility for generative and stylization settings.

\subsection{DiffVG and applications}
In recent years, differentiable rendering has enabled the use of large pretrained vision and generative imaging models with 3D \cite{katoDifferentiableRenderingSurvey2020,tewariStateArtNeural2020,Worchel2023}  and 2D \cite{mihaiDifferentiableDrawingSketching2021,liDifferentiableVectorGraphics2020,Worchel2023} parametric primitives \revnote{Simply added ref to Worchel and Alexa think discussion is sufficient}. We adopt the method of \citet{liDifferentiableVectorGraphics2020}, which supports a large subset of the SVG standard and cubic curves with varying width profiles. Our method leverages DiffVG's support for cubics with varying width profiles, a feature
yet to be used comprehensively, likely due to limited support in mainstream vector graphics tools and standards.

\paragraph{CLIP-driven graphics.}
One of the first applications of DiffVG to large-pretrained models has been through the
use of the Contrastive Language–Image Pretraining (CLIP)
model \cite{radfordLearningTransferableVisual2021}, a multimodal model that has
been trained to share an embedding space between images and their textual
descriptions. \citet{FransClipdraw2022} demonstrate that together with DiffVG, the model is able to generate vector images guided by a text caption or ``prompt''. 
\citet{ganzCLIPAGGeneratorFreeTexttoImage2024} use an adversarial "robustification" method to fine-tune CLIP in order to enable gradients that are better aligned with human perception. \citet{vinkerCLIPassoSemanticallyAwareObject2022} introduce the idea of using a loss on internal layers of CLIP to guide vector image abstraction. A similar approach, combined with DiffVG,
has enabled the generation of stroke-based stylization methods \cite{xingDiffSketcherTextGuided2024,vinkerCLIPasceneSceneSketching2023,schaldenbrandFRIDACollaborativeRobot2022}. Our method provides similar capacities, but we take advantage of the fine-tuned CLIPAG model of \citet{ganzCLIPAGGeneratorFreeTexttoImage2024} and support long smooth strokes, which was not possible with previous methods.

\paragraph{Diffusion-driven graphics.}
 In the context of 3D asset generation, \citet{poole2023dreamfusion} pioneered the so-called Score Distillation Sampling (SDS), which enables gradient propagation from pre-trained diffusion models to parametric representations. While effective, the original method relies on high classifier-free guidance (CFG) scales, often resulting in over-saturation and lack of detail \cite{katzir2024noisefree}. Recent methods, including variational methods \cite{WangProlific2023}, DDIM inversion \cite{liangLucidDreamerHighFidelityTextto3D2023} and noise-free score distillation (NFSD) \cite{katzir2024noisefree}, address these challenges, improving fidelity and control. Our method is compatible with all these techniques, and we specifically adopt the approach of \citet{liangLucidDreamerHighFidelityTextto3D2023}, which proved to be the most effective for us. \deleted{We demonstrate how these diffusion model gradients can be used in our differentiable rasterizer to produce image abstractions. }

In the context of 2D asset generation, \citet{jainVectorFusionTexttoSVGAbstracting2022} pioneered the use of SDS in conjunction with the DiffVG method of  \citet{liDifferentiableVectorGraphics2020}, demonstrating the expressive potential of diffusion for vector graphics generation. \citet{iluzWordAsImageSemanticTypography2023} use SDS for stylizing vector font outlines to resemble user-defined semantics. To name a few, variants of SDS have been used for prompt-based sketch generation \cite{xingDiffSketcherTextGuided2024} and animation \cite{Gal_2024_CVPR},  2D vector graphics \cite{xingSVGDreamerTextGuided2024,zhangTexttoVectorGenerationNeural2024} as well as 3D line art \cite{tojoFabricable3DWire2024,quWiredPerspectivesMultiView2024}.
None of these methods support the creation of long, smooth strokes aligned with our objectives, except for \citet{tojoFabricable3DWire2024}, who also use B-splines 
to produce long single-stroke outputs. We incorporate their proposed repulsion loss in our method. However, their work does not cover high-order derivative smoothing, relies on curve discretization, uses a custom CUDA renderer, and does not support variable-width strokes.

\begin{figure}[t] %
	\centering
	\includegraphics[width=0.49\textwidth]{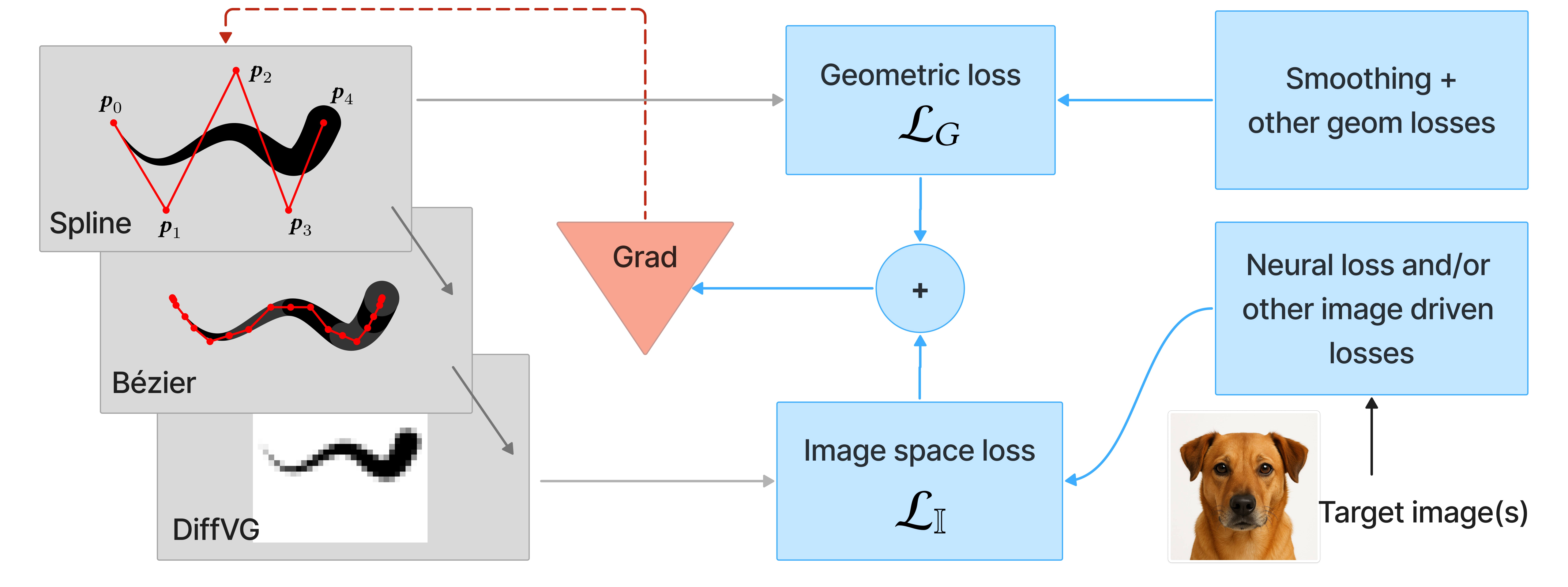}
	\caption{
      Flow chart of our pipeline, all operations are differentiable.
  } %
\label{fig:process}
\end{figure}

\begin{figure}[t]
  \centering
  \includegraphics[width=0.45\textwidth]{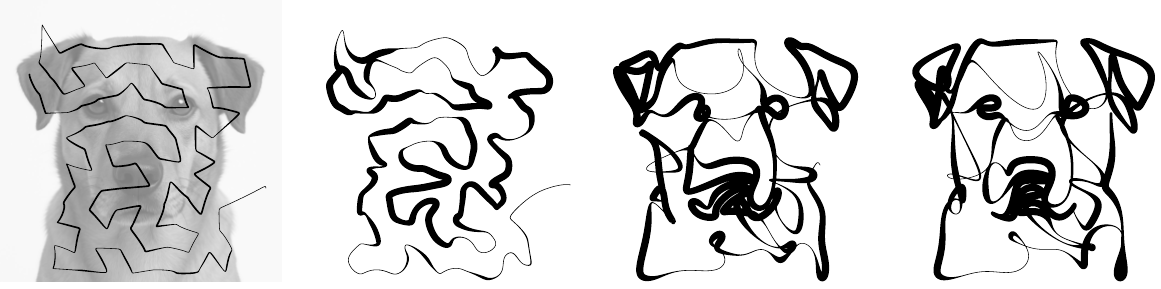}
	\caption{
      Optimization procedure. From left to right: input image with an initial spline (quintic with multiplicity $3$ on all keypoints) and subsequent optimization steps $10, 150, 300$.
    } %
\label{fig:ism-steps}
\end{figure}

\section{Method}

Our approach works as follows: we specify one or more B-splines through a series of 2D or 3D \textit{keypoints}, where the third dimension can be used to describe width variation along a stroke. The B-splines are converted to cubic piecewise B\'ezier curves that are then rendered in a differentiable manner with the method of ~\citet{liDifferentiableVectorGraphics2020} (see also \autoref{fig:process}). Similarly to conventional DiffVG pipelines, this enables gradient optimization of the key-points with costs that depend on curve geometry as well as 
on the rendered version of the curves. Figure \ref{fig:ism-steps} shows the process: first, some initial points and an image are given; then, during optimization a spline gradually represents the input image more explicitly, while the stroke widths are jointly adapted.

\subsection{Uniform B-splines}
We use normalized uniform or ``cardinal'' B-splines
that have uniformly spaced integer knots \cite{De_Boor2001-bd}, which simplifies computations and proves successful in our applications. A B-spline
of degree $p$ and order $k=p+1$ is a linear combination
\[
\bm{x}(u) = \sum_{i=0}^{n-1}{\cp_i N_{k}(u - t_i)}
\]
where $n$ control-points \(\bm{C}=\left[\cp_0, \cp_1, \dots
\cp_{n-1}\right]\) and shifted bases $N_{k}$ are associated with a non-decreasing sequence of \(m=n+k\) knots. In our formulation we keep these fixed to
\[
\bm{t}=[t_0, \dots, t_{m-1}]=[-p, \dots,
\underbrace{0, \dots, n-k}_{t_p, \dots, t_{m-k}}, \dots n].
\]
The spline is defined by sampling $u$ in the interval \(\left[t_{k-1}, t_{m-k}\right]\). Increasing order derivatives $\bm{x}^{(d)}$ of a B-spline are easily computed as weighted combinations of lower order B-splines. We refer the reader to the supplement for details on the basis functions construction, but these are readily available in many modern scientific computation packages \cite{2020SciPy-NMeth}. The number of curves and control points is predefined, so that basis functions and knot sequences can be precomputed and remain fixed during optimization.

\subsection{Spline construction}
\label{sect:spline_constr}

B-splines are approximating curves, and both periodicity and clamping to endpoints require the repetition of either knots or control points. This is typically achieved with repeated knots, but we follow \citet{fujiokaConstructingReconstructingCharacters2006} and use repeated control points. 
This maintains strict uniformity while enabling adaptive smoothing of corner-like features and simplifying integral computations, which is advantageous for our use-case. 
Instead of directly specifying control points, we let a user initially specify a spline through a series of $M$ \textit{key-points} \(\bm{Q}=\bm{q}_1, \dots, \bm{q}_{M}\) and \textit{optimize these rather than the spline control points directly}. The key-points are automatically adapted into a series of control points $\bm{C}$ depending on the curve's desired clamped or periodic behavior.

For a clamped (open) spline the control points are given by the key-points $\bm{Q}$ padded the first and last key-point repeated $k-1$ times. This results in a parametric motion that begins and ends with a rest.
For periodic closed splines we construct $\bm{C}$ by appending the first $k-1$ keypoints to the initially specified key-point sequence $\bm{Q}$.  %

Key-points may optionally be repeated to create sharp corners, as each repetition initially reduces the continuity of the curve by one degree \cite{farinCurvesSurfacesCAGD2001}. This strategy is useful to produce additional degrees of freedom for the subsequent optimization, where the corners can be adaptively smoothed depending on the desired amount of smoothing.

\subsection{Smoothing B-splines}
\label{sect:smoothing}
\begin{figure}[b]
	\centering
	\includegraphics[width=0.49\textwidth]{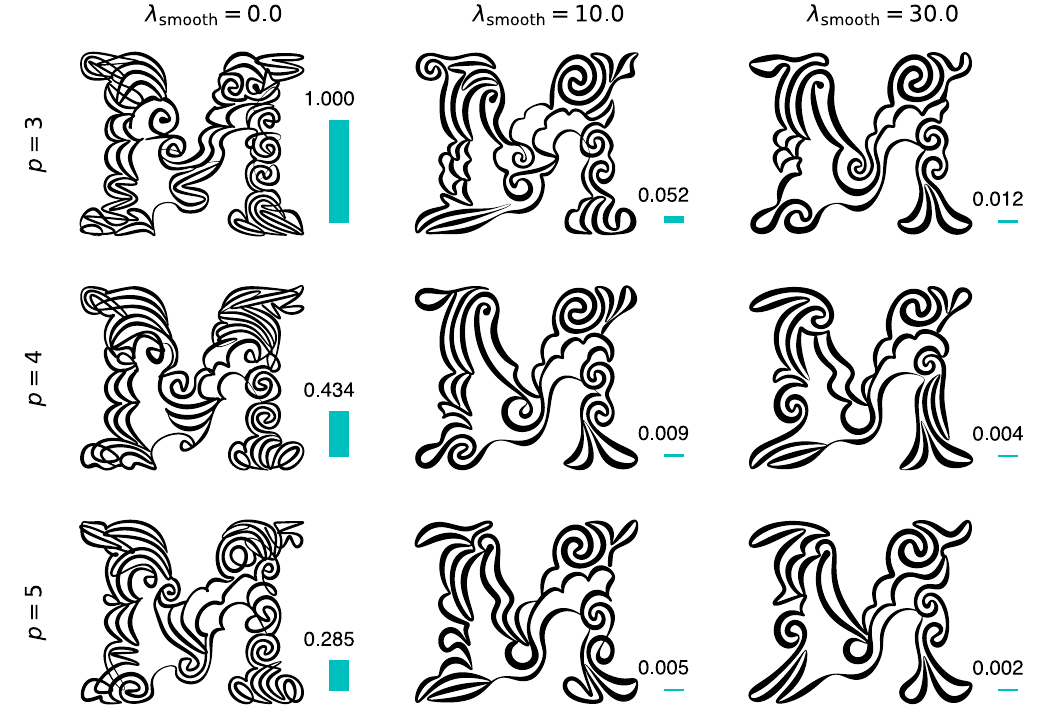}
	\caption{
    \revision{Comparison of different spline degrees $p$ (rows), smoothing derivative orders $d$ and smoothing weight $\lambda_{\text{smooth}}$ (columns). In each row, we let the smooting derivative to $p-1$. We quantify smoothness using the dimensionless jerk measure \cite{Hogan2009dim}. Lower is smoother. We use the stylized area fill method in \refsect{sec:AreaFilling} using the style image in \refig{fig:more-s}, left.}
    } %
\label{fig:abl}
\end{figure}

B-splines of order $k$ are by definition $C^{k-2}$-continuous, but more importantly their construction facilitates the formulation of smoothing criteria since they allow closed form computation of derivatives and integrals. In our method, we adopt a smoothing cost based on the squared magnitude of the curve derivatives, which is standard in the smoothing literature and is also known for its utility in curve fairing \cite{pottmannSmoothCurvesTension1990} and for modeling human arm movements \cite{todorovSmoothnessMaximizationPredefined1998}. These methods typically trade off smoothness with a geometric accuracy term, but in our work we consider a variety of image-space objectives instead of geometry and define a smoothing cost:
\begin{equation}
  \mathcal{L}_{\mathrm{smooth}}^{d} = \frac{1}{T} \int_{t_{k-1}}^{t_{m-k}} { \|\bm{x}^{(d)}(u)\|^2 \mathrm{d}u } = \frac{1}{T} \bm{c}^\trsp \bar{\bm{G}} \bm{c}
\end{equation}
where $T=t_{m-k}-t_{k-1}$ and $\bm{c}$ is a vector that concatenates all control points of the spline. The integral can be calculated exactly by setting $\bar{\bm{G}}$ to a block Gram-matrix constructed from the inner products of the basis function derivatives \cite{vermeulen1992integrating, fujiokaConstructingCharacterFont2007}, resulting in the standard spline smoothing criterion. Alternatively, a finite difference approximation of $\bar{\bm{G}}$ results in the penalized-spline method of \citet{eilersFlexibleSmoothingBsplines1996}. Both methods have similar run-time performance because the matrix is precomputed for each stroke, and we refer the reader to the supplement for derivations. 
Most of our examples use quintic splines with a smoothing cost $\mathcal{L}_{\mathrm{smooth}}^{3}$ on the third positional derivative (jerk). We do so on the basis that ``minimum jerk'' is a known criterion that has been used to model hand and arm movements \cite{todorovSmoothnessMaximizationPredefined1998,flashCoordinationArmMovements1985} as well as an \correction{good approximation}{approximant} for curvature variation in curve fairing \cite{luNoteCurvatureVariation2015}. Nevertheless, our method generalizes to different curve and smoothing orders \revision{(\refig{fig:abl})}.

\subsection{Conversion to B\'ezier and rendering}
Our goal is to integrate smoothing B-splines into a DiffVG pipeline by taking advantage of the linear relationship between B-splines and B\'ezier curves.
B-splines can be converted exactly to piecewise B\'ezier curves of the same degree. To do so we use the method of \citet{romaniConversionMatrixUniform2004a}, which reduces to a matrix multiplication between the flattened spline control points $\bm{c}$ and a block transformation matrix $\bm{S}$.

Our method also supports smoothing costs on higher-order positional derivatives such as jerk (third derivative) and snap (fourth derivative), which require polynomial curves of degree greater than three.
Although native rendering of such higher-degree curves is not supported in DiffVG and remains a challenge, we observe that reducing the degree of B-splines to three introduces negligible geometric error (less than \(0.3\%\) of the curve's bounding box diagonal in all our experiments), making the optimization of higher-degree B-splines practical for image-based error calculations.

We perform a degree reduction of B\'ezier curves using the multi-reduction method of \citet{sunwooMatrixRepresentationMultidegree2005}, which involves a second block transformation matrix $\bar{\bm{R}}$. As a result, the control points for a cubic piecewise Bézier curve compatible with DiffVG are computed from the (flattened) control points $\bm{c}$ with the linear map $\bar{\bm{R}}\bar{\bm{S}}\bm{c}$. We refer the reader to the work of \citet{romaniConversionMatrixUniform2004a} and \citet{sunwooMatrixRepresentationMultidegree2005} for details; we include in the supplement details and matrices for quintic B\'ezier and their reduction to cubic.

\paragraph{DiffVG rendering and optimization}
The conversion procedure results in a sequence of \revision{B\'ezier control points %
  $\in \reals^3$
}, where the third dimension represents the stroke radius.
Control points and associated stroke and fill colors are all treated as differentiable parameters to be optimized.
Rendering the scene results in an image $\mathbb{I}$, which is differentiable with respect to all the underlying parameters.

\section{Applications}
\label{sect:app}

The proposed B-spline construction, smoothing and conversion to B\'ezier enables the optimization of long, expressive and optionally periodic curves,
which would be challenging to produce with currently known methods leveraging DiffVG. 
All the results presented hereafter are produced using a combined cost:
\begin{equation}
  \mathcal{L} = \mathcal{L}_{\mathbb{I}} + \mathcal{L}_{\textrm{G}}
  \label{loss:total}
\end{equation}
\noindent consisting of an image-space term, \(\mathcal{L}_{\mathbb{I}}\), and a geometric term, \(\mathcal{L}_{\text{G}}\).
We construct each term as a combination of losses depending on the application objective.
\(\mathcal{L}_{\mathbb{I}}\) relies on differentiable
rasterization, which allows gradients to propagate from raster-based objectives to the geometry parameters.
\(\mathcal{L}_{\text{G}}\) leverages the properties of B-splines to enable
smoothing, stylization objectives, and constraints while preserving continuity.
We denote the relative weights of any loss $\mathcal{L}_{\circ}$ as $\lambda_{\circ}$, e.g. the weight of a smoothing loss on the third derivative is denoted as $\lambda_{\text{smooth}}$. If not specified, the weights are assumed to be $1$. When also optimizing stroke widths, we clip these to a minimum and maximum value at each iteration.

We generate strokes using the Adam optimizer and use a cosine annealing schedule on the learning rates. We run our experiments on a single NVIDIA GeForce RTX 3060 with 12 Gb of memory. We run most of the presented applications for $300$ steps, which approximately takes between $30$ and $60$ seconds on our system. One exception is using diffusion-guidance, which
takes approximately $0.6$ to $1.0$ second per step depending on the method used, leading to an optimization time of up to $6$ minutes. 
\deleted{All the geometric losses we present are linear in the number of key-points, with the exception of the repulsion loss of 
which requires sampling and is quadratic in the number of samples.}

\begin{figure}
	\centering	\includegraphics[width=0.44\textwidth]{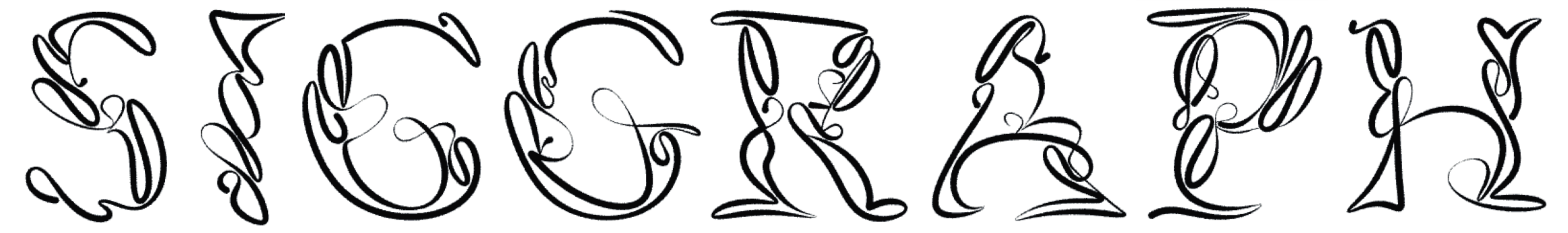}
	\caption{
      Text combining areas generated with our stylized area filling method. Each letter is generated separately.
} %
\label{fig:siggraph-area}
\end{figure}
\begin{figure} %
	\centering
	\includegraphics[width=0.47\textwidth]{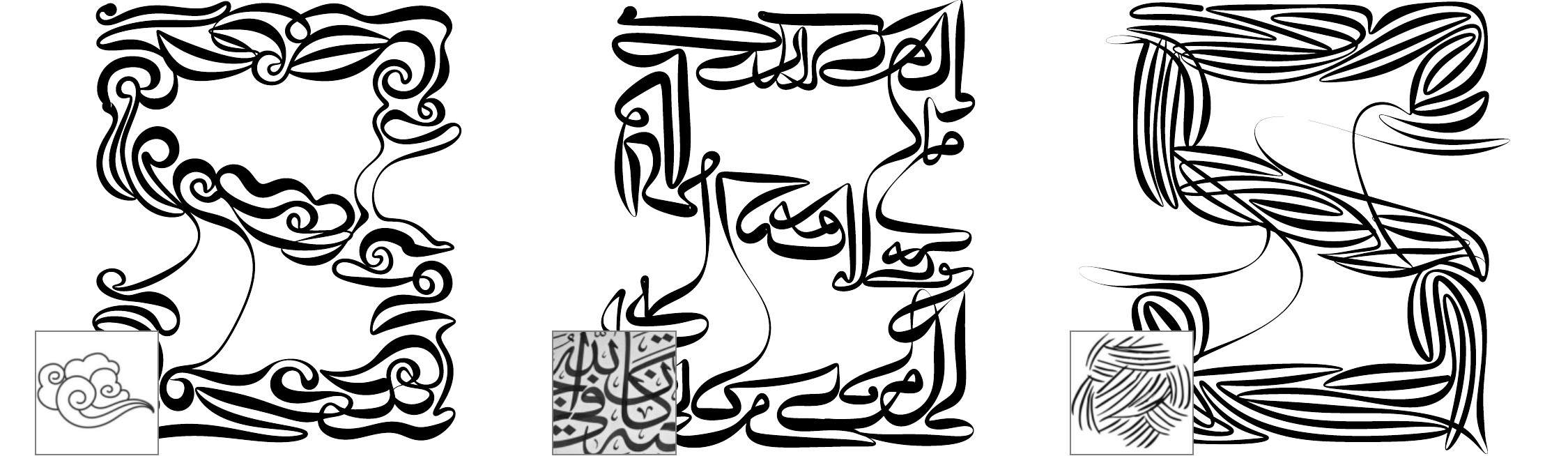}
	\caption{
\revision{      Examples of stylized area filling for a letter ``S''. The images on the lower left are used to guide stylization.}
} %
\label{fig:more-s}
\end{figure}

\subsection{Area fillings and pattern generation}
\label{sec:AreaFilling}
As a baseline for our method we demonstrate how our pipeline can be used to create pattern fills of solid regions. It illustrates also how our approach can be flexibly used to control stylization while maintaining smoothness \revision{(\refig{fig:siggraph-area} and \ref{fig:more-s})}.

\begin{figure}[t]
	\centering
	\includegraphics[width=0.47\textwidth]{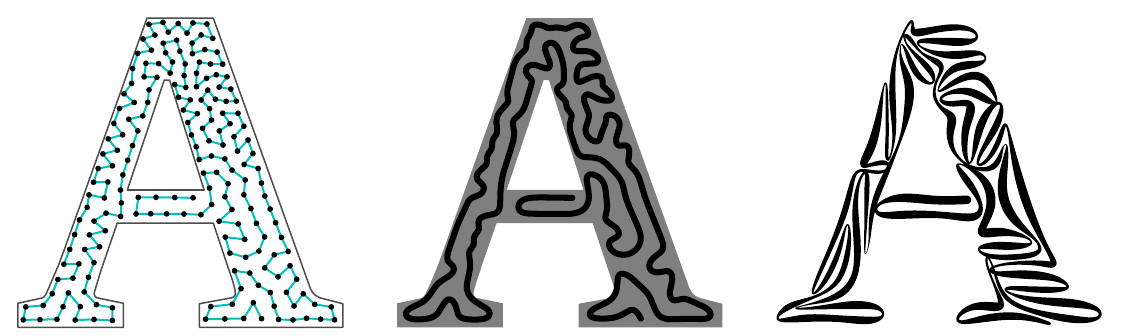}
	\caption{
      Left, weighted Voronoi samples (black) for a bitmap of the letter 'A' and an open TSP path connecting the points (blue).
      Middle, initial quintic B-spline with key-points given by the Voronoi samples overlaid on the bitmap area with $50\%$ opacity. Right, result of an optimization with $\mathcal{L}_{\text{G}}=\mathcal{L}_{\text{smooth}}^{3}$ and $\mathcal{L}_{\mathbb{I}}$ for the $50\%$ opacity bitmap. Decreasing opacity results in sparser and thinner strokes.
    } %
\label{fig:init}
\end{figure}

\paragraph{Initialization}
Stochastic gradient descent is well known to be sensitive to initialization due to its susceptibility to local minima. We find good points using an initialization strategy based on so-called weighted Voronoi stippling~\cite{secordWeightedVoronoiStippling2002}. 
For simplicity, we adopt this method for different applications presented in this paper. 
The input can be an arbitrary bitmap (\refig{fig:init}, Left) or a saliency map (\refig{fig:spock}).
To create a single stroke, we use a TSP route connecting the points in an open or looping path. This method is known in the literature as ``TSP art''  \cite{kaplan2005tsp} (\refig{fig:init}, Left).
For open paths, we select the left-topmost point and the bottom-rightmost as initial and final points, respectively.

\paragraph{Image coverage loss}
\label{sect:coverage}
We find that setting $\mathcal{L}_{\mathbb{I}}$ as a multiscale mean squared error (MSE) loss works particularly well to fill an area or silhouette defined as an image. This loss is computed between the target and the rendered image, with each step corresponding to a progressively reduced scale and blurred version of the image. This approach is similar to the shape-based losses used by \citet{iluzWordAsImageSemanticTypography2023} and \citet{tojoFabricable3DWire2024}, but lower scales encourage alignment with broader intensity regions and faster convergence, while higher scales promote a more accurate silhouette reconstruction. 
Reducing the opacity of the target image directly decreases the density of curves used to cover it (\refig{fig:init}-right), allowing control over the visual result.

\begin{figure}[t]
	\centering
	\includegraphics[width=0.46\textwidth]{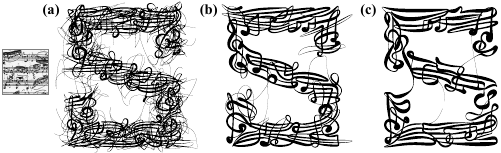}
	\caption{
    Stylized coverage of a letter ``S'' using the image on the left as a target. \textbf{(a)} Optimization using B\'ezier curves. \textbf{(b)} Optimization using quintic splines and smoothing on jerk with $\lambda_{\text{smooth}}=1$. \textbf{(c)}. Same procedure with $\lambda_{\text{smooth}}=10$.
} %
\label{fig:smooth-comp}
\end{figure}

\paragraph{Bounding box loss} For some of our optimization procedures, it is useful to extend \(\mathcal{L}_{\text{G}}\) with a bounding box loss that keeps curve key-points within the bounding box of a given image:
\[
\mathcal{L}_{\text{box}} = \sum_i
\bm{1}^\trsp \left[
    \varphi\left(\bm{b}_{\min} - \bm{p}_i\right) +
    \varphi\left(\bm{p}_i - \bm{b}_{\max}\right)
\right]
\]
where \(\mathcal{L}_{\text{box}}>0\) only if key-points fall outside of the bounding box $\bm{b}_{\min}, \bm{b}_{\max}$ %
and where $\varphi$ can be either a $\text{Softplus}$ or a $\text{ReLU}$ function applied element-wise to the vectors.

\begin{figure}[t]
	\centering
	\includegraphics[width=0.47\textwidth]{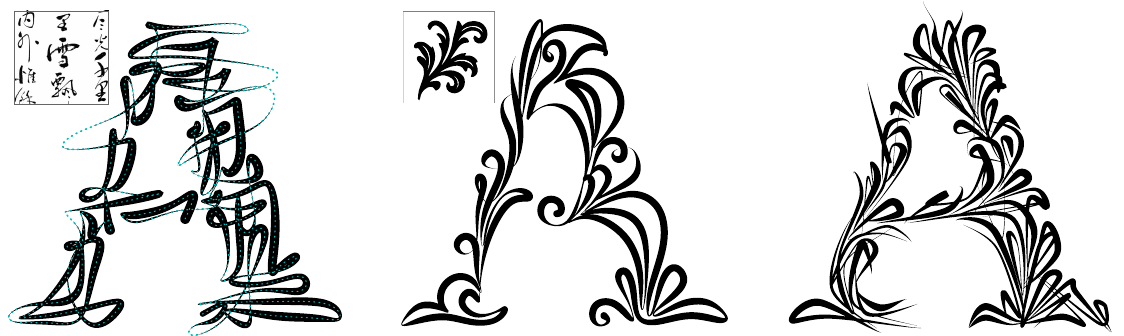}
    \caption{
      Examples combining image coverage with a patch-wise loss on CLIP features derived from an example image (top left) and using the same initialization from \refig{fig:init}. From the left, the first two examples use a quintic spline with a smoothing loss $\mathcal{L}_{\text{G}}=\mathcal{L}_{\text{smooth}}^{3}$ (jerk). For comparison, the right example uses a Catmull-Rom spline only enforcing $C^{1}$ continuity. Allowing zero stroke width results in the appearance of multiple strokes, but the optimization is still performed on a single curve (left, dotted cyan). %
} %
\label{fig:clip-patch}
\end{figure}

\paragraph{Image-space semantic-driven stylization}
Together with geo\-metry-based stylization costs, we can add a semantic stylization term $\mathcal{L}_{\text{style}}$ to the image-space loss $\mathcal{L}_{\mathbb{I}}$, which enables stylization based on a text prompt or features extracted from an example image (\revision{ \refig{fig:smooth-comp} , \ref{fig:more-s} and \ref{fig:clip-patch}}).
We apply the technique proposed by \citet{kwonCLIPstylerImageStyle2022} for seman\-tic-driven image stylization and use a patch-wise directional loss between the encoded features of an example image and the encoded features of the rendered curves.
We use the augmented CLIPAG \cite{ganzCLIPAGGeneratorFreeTexttoImage2024} ViT-B/32 transformer architecture as we find it to be efficient while working well for our use vector stylization use-case.

\subsection{Single-stroke image abstraction}
\label{sec:ImageAbstraction}

\begin{figure}[t]
	\centering
	\includegraphics[width=0.45\textwidth]{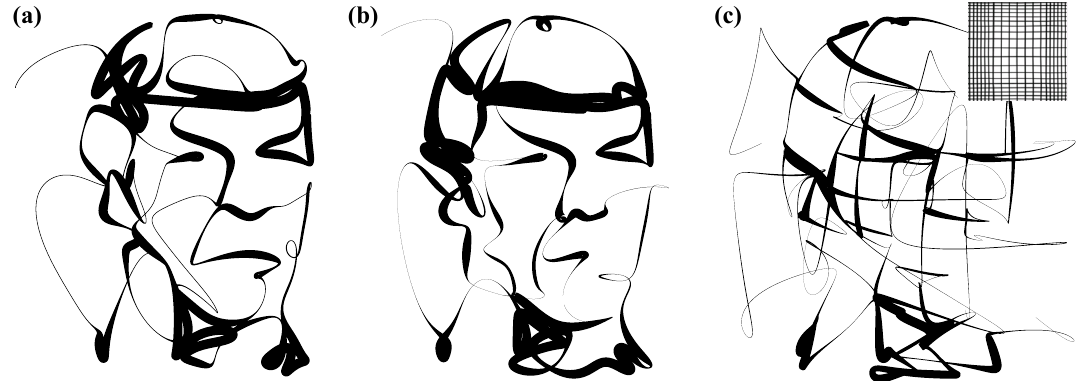}
	\caption{
      Diffusion driven stroke abstraction with ControlNet and IP-adapter conditioning. \textbf{(a)} a variable width abstraction of ``Spock''. \textbf{(b)} Allowing a single strokes to reach zero width in regions results in an effective strategy for automatically determining the number of strokes for a multi-stroke abstraction. \textbf{(c)} combining the diffusion cost with a stylization term \revision{that favors horizontal and vertical orientations}. %
} %
\label{fig:sds-stroke}
\end{figure}

\begin{figure}[t]
	\centering
	\includegraphics[width=0.45\textwidth]{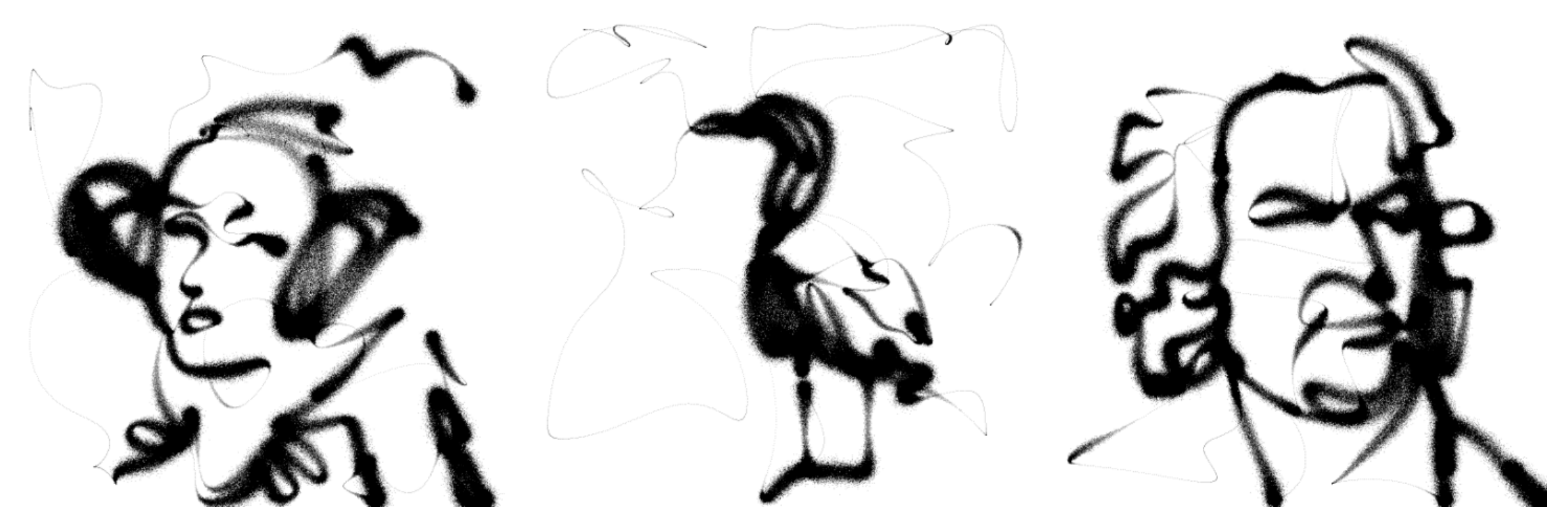}
	\caption{
      Rendering stroke abstractions with a spray-like brush. The quintic B-splines together with smoothing on jerk produce smooth motions that tend to slow down where curvature is higher. This is a characteristic feature of human hand motions \cite{vivianipaoloandflashtamarMinimumJerkTwoThirdPower1995} and results in a lower deposition of paint particles where speed is higher.
} %
\label{fig:spray}
\end{figure}

Most existing DiffVG-based methods that work with diffusion models rely on variants of Score Distillation Sampling (SDS) together with a text caption to guide the generation of parametric vector primitives. We follow a similar approach but enable image-conditi\-oned stylization by integrating ControlNet \cite{zhangAddingConditionalControl2023} with Canny edge detection and IP-Adapter \cite{ye2023ipadaptertextcompatibleimage} into the diffusion pipeline. ControlNet helps to preserve structural cues from the input image, while IP-Adapter encourages the strokes to align with its global appearance and style \revision{(\refig{fig:sds-stroke} and \ref{fig:spray})}.

In our experiments, we find that using a generic text prompt such as ``A black and white drawing'' for stroke-based outputs is sufficient to generate recognizable abstractions and stylizations of an input image. For a given condition \(y\), the gradient of the SDS-like loss with respect to the optimizated parameters \(\theta\) has the form:
\begin{equation}
  \bm{\nabla}_\theta \mathcal{L}_{\textrm{SDS}} = \mathbb{E}_{t} \left[ \omega(t) \left( \bm{\epsilon}_\phi(x_t, t, y) - \bm{\epsilon} \right) \frac{\partial g(\theta)}{\partial \theta} \right],
  \label{eq:ism}
\end{equation}
\noindent where \(\bm{\epsilon}_\phi(x_t, t, y)\) is the predicted denoising direction for a latent \(x_{t}\) at time step \(t\), \(\bm{\epsilon}\) is the noise predicted by the model and \(\omega(t)\) is a weighting 
function dependent on the time-step.

We employ the time-step schedule annealing procedure proposed by
\citet{liangLucidDreamerHighFidelityTextto3D2023}
and use their Interval Score Matching (ISM) variant of SDS, %
which helps convergence in our experiments and enables a standard classifier-free guidance of $7.5$.

\begin{figure}[t]
	\centering
	\includegraphics[width=0.46\textwidth]{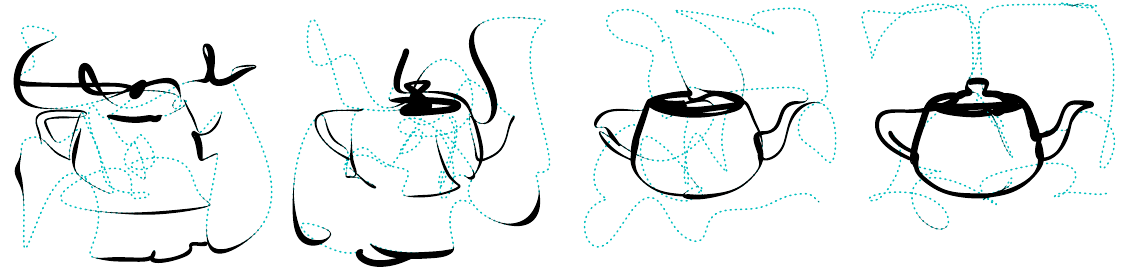}
	\caption{
    From left to right: varying the minimum time-step ($100, 300, 500, 700$) for $300$ timesteps of the ISM \cite{liangLucidDreamerHighFidelityTextto3D2023} variant of SDS and a single quintic stroke. The cyan line emphasizes the centerline of the stroke, which reaches zero width in certain regions.
} %
\label{fig:sds-utah}
\end{figure}

\begin{figure}[t]
\includegraphics[width=0.47\textwidth]{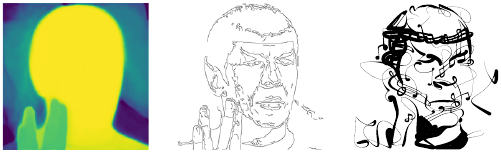}
	\caption{
      \revision{Left, initialization with a saliency map computed from the normalized logits of the last layer of the OneFormer panoptic segmentation model~\cite{jainOneFormerOneTransformer2023}.
  Right, stroke optimization using ISM with diffusion conditioned on the edge map, using a minimum time step of $400$ and with an additional stylization loss $\lambda_{\text{style}}$ guided by the same image as \refig{fig:smooth-comp}.}
} %
\label{fig:spock}
\end{figure}

It is known that for diffusion models, higher time steps during denoising typically produce coarser features, while lower time steps yield finer details \cite{hwangResolutionChromatographyDiffusion2023}.
Given our goal of producing single stroke image abstractions, the curves lack sufficient degrees of freedom to capture these finer details, so we limit the time steps in the denoising process to a minimum of $500$ \revision{(\refig{fig:sds-utah})}.
With a similar motivation, we find that with diffusion-guided stroke abstraction it is useful to initialize the strokes with a multiplicity $>1$ (we use $3$ with quintic splines in our examples). Using a higher multiplicity results in smoother strokes, where fewer details are captured.

\subsection{Area-based image abstraction }
\label{sec:AreaAbstraction}

Our method allows for the generation of smooth closed areas, and we observe that this is useful to generate image abstractions similar to what can be seen in certain designs consisting of overlapping smooth regions and a limited color palette. Examples include psych-edelic %
designs, album covers, screen-printed graphics, or street-art inspired fashion and graphic designs. We are interested in generating outputs that aim to be printed or fabricated as collages with a limited number of regions and colors. To guide stylization, we use filled areas instead of strokes and set $\mathcal{L}_{\img}$ to a variant of the CLIP-driven geometric cost described by \citet{vinkerCLIPassoSemanticallyAwareObject2022}
\begin{equation}
  \mathcal{L}_{\text{CLIP}} = \sum_l \left\| \text{CLIP}_l(\hat{\mathbb{I}}) - \text{CLIP}_l(\mathbb{I}_{\theta}) \right\|_1,
\label{eq:clip}
\end{equation}
using the $L^{1}$ norm instead of $L^{2}$ and omitting the semantic term originally proposed by the authors. We use layers \(2\) and \(3\) together with the \deleted{augmented} CLIPAG \cite{ganzCLIPAGGeneratorFreeTexttoImage2024} architecture. We use CLIP as opposed to diffusion because we find this to be significantly faster, while being effective for this kind of stylization task.

\begin{figure}[t]
	\centering
	\includegraphics[width=0.49\textwidth]{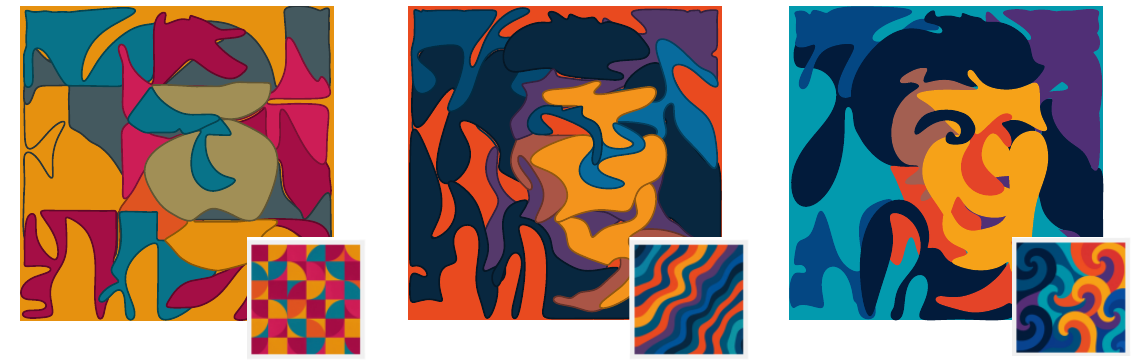}
	\caption{
      Quantized color vectorizations using an additional image-driven stylization term  $\mathcal{L}_{\text{style}}$. The palette is extracted from the style image.
} %
\label{fig:vector-style}
\end{figure}

\paragraph{Repulsion loss}
For applications using closed curves, we adopt the repulsion method for 3D wire fabrication proposed by \citet{tojoFabricable3DWire2024} to compute a geometric loss $\mathcal{L}_{\text{repul}}$, penalizing self-intersections and overlaps based on a tangent-point energy kernel for a set of sampled points along the spline. For this application, we compute the loss for each area separately, thus allowing overlaps and intersections among different areas.

\paragraph{Optimization with quantized coloring}

\citet{jang2017categoricalreparameterizationgumbelsoftmax} use the Gumbel-Softmax trick to make discrete choices differentiable during training. We apply the same idea to assign colors to image regions, using soft selections from a fixed color palette that can be optimized with backpropagation. To progressively transit from soft to discrete assignments during the training process, we anneal the Gumbel-Softmax temperature using an exponential schedule.

\begin{figure}[t]
	\centering
	\includegraphics[width=0.48\textwidth]{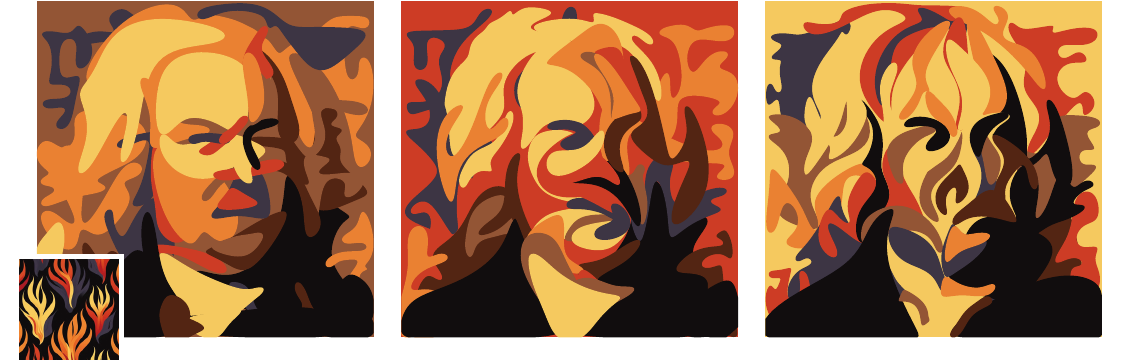}
	\caption{
      Increasing $\lambda_{\text{style}}$ weight for an abstract vectorization of Bach. From left to right $\lambda_{\text{style}}=0$, $\lambda_{\text{style}}=0.06$ , $\lambda_{\text{style}}=0.1$
} %
\label{fig:bachfire}
\end{figure}

Given a set of $K$ palette colors organized as a matrix \revision{$\bm{V} \in \reals^{K\times3}$},
we optimize the logits per area $\ell_i \in \mathbb{R}^K$ using a soft assignment
\[
  \bm{a}_i = \text{softmax}\left( \frac{\ell_i + \mathbf{g}_i}{\tau} \right)
  \quad\text{with}\quad
  \bm{g}_i \sim \text{Gumbel}(0, \beta)^K,
\]
where $\beta$ is a scale parameter that we empirically set to $0.15$ to avoid excessive noise during
optimization \cite{huijbenReviewGumbelmaxTrick2022} and $\tau$ is a temperature parameter that we anneal during optimization. We use these soft colors computed as \revision{$\mathbf{v}_i = \bm{a}_i^\trsp \bm{V}$} during the optimization. At the same time, for visualization, we obtain hard assignments by taking the argmax over the optimized logits and selecting the corresponding palette color. To encourage a balanced use of all the specified palette colors, we add a regularization term:
\[
  \lambda_{K} \left\| \mathbb{E}_{i} \left[ \bm{a}_i \right] - K^{-1}\bm{1} \right\|^2
\]
to $\mathcal{L}_{\img}$, which penalizes deviations from a uniform color assignment, encouraging a balanced use of the palette. Figures \ref{fig:vector-style} and \ref{fig:bachfire} show some results.

\paragraph{Area initialization and optimization} We initialize a user-defined number of areas using weighted Voronoi sampling on a saliency map of the input image and create an initial series of closed curves with keypoints given by the vertices of each resulting Voronoi regions. Each curve is then assigned random initial logit and the curves are sorted by increasing saliency of the covered area. Optimization proceeds with the inclusion of the repulsion loss in \(\mathcal{L}_{\text{G}}\), which keeps the area outlines from intersecting.
\subsection{Text stylization}
\label{sec:TextStylization}

\begin{figure}[b] %
	\centering
	\includegraphics[width=0.49\textwidth]{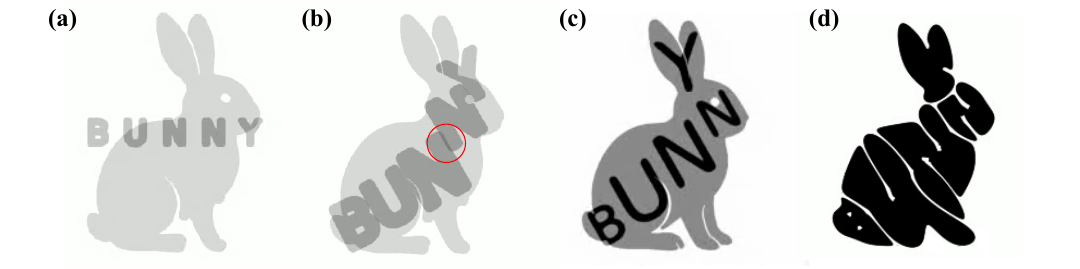}
	\caption{
      Automatic calligram production for a silhouette generated with the prompt ``Silhouette of a BUNNY``. \textbf{(a)} initial text layout rendered with 50\% opacity and overlayed on the silhouette. \textbf{(b)} intermediate step of the layout optimization displaying an image area that increases the overlap cost. \textbf{(c)} Sampled glyphs placed according to the layout. \textbf{(d)} Result of the optimization.
} %
\label{fig:gpt-bunny}
\end{figure}

\begin{figure}[t]
	\centering
	\includegraphics[width=0.45\textwidth]{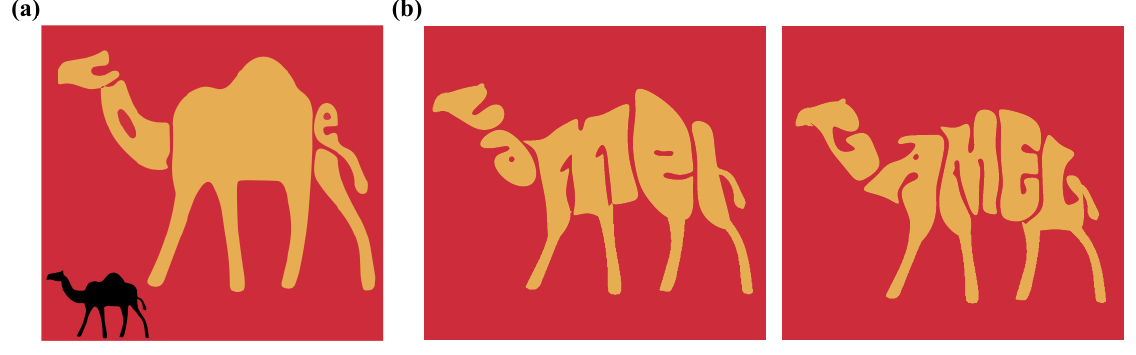}
	\caption{
      Calligram generation: comparison of \textbf{(a)} an \revision{example from \citet{zouLegibleCompactCalligrams2016} for a camel silhouette} and \textbf{(b)} two runs of our method \revision{on the same silhouette} with automatic initialization and two different fonts.
} %
\label{fig:calligram-comp}
\end{figure}

\begin{figure}[t]
	\centering
	\includegraphics[width=0.47\textwidth]{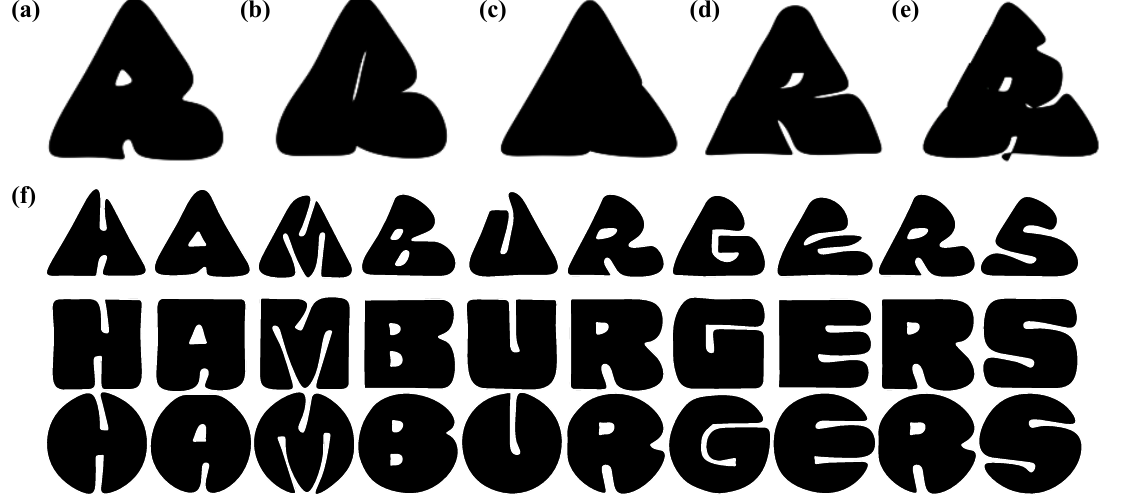}
	\caption{
    Monospace font generation. \textbf{(a)} A letter ``R'' (quintic B-splines with jerk cost) adapted to a triangle, using $\lambda_{\text{repul}}=666$, $\lambda_{\text{txt}}=6.6$ and $\lambda_{\text{smooth}}=200.0$. \textbf{(b)} Setting $\lambda_{\text{repul}}=0$ (no repulsion), still results in a readable letter but \textbf{(c)} removing the legibility loss does not.
        \textbf({d}) B-splines with legibility but no smoothing. \textbf{(e)} Catmull-Rom to enforce tangent continuity with legibility loss (for comparison). \textbf{(f)}
      Combining glyphs optimized to fit a triangle, a square and a circle.
} %
\label{fig:hamburger}
\end{figure}

In line with the smooth curve image abstractions, we aim to generate text abstractions made of smooth curves that fit inside a target area. Examples of this approach can be seen in posters, graphic designs, as well as in ``calligrams'': renditions of text that is arranged to fit a specific silhouette, such as those seen in the methods of \citet{xuCalligraphicPacking2007} and \citet{zouLegibleCompactCalligrams2016} \addition{(c.f. \autoref{fig:calligram-comp})}. Our pipeline results in a simple way to generate calligrams, such as ``blobby'' texts (\refig{fig:gpt-bunny}) and abstract monospace fonts (\refig{fig:hamburger}).

We tackle text stylization with the tools we have covered so far and start with a bitmap image $\hat{\img}$ representing the desired silhouette and an initial text layout rendered as a second image $\img_{\text{txt}}$. We uniformly sample the glyph outlines and produce key-point sequences used in optimization. The optimization deforms the outlines based on a loss that balances silhouette coverage, outline smoothness, and repulsion between outline points. This procedure alone smooths and fits the outlines into the target area, but this may compromise legibility (\autoref{fig:hamburger}c).


\begin{figure}[t]
	\centering	\includegraphics[width=0.45\textwidth]{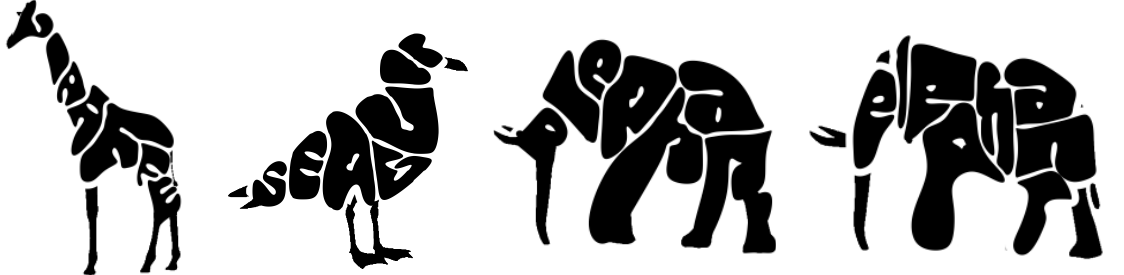}
	\caption{
      More calligrams generated with our system. \revision{The seagull silhouette is generated using the prompt ``Silhouette of a SEAGULL''.}
} %
\label{fig:more-calli}
\end{figure}

To preserve \textit{legibility}, we introduce a perceptual loss based on the features of a pretrained vision encoder, which we use to compute the feature-space distance between the rendered deformed image $\img$ and the original layout $\img_{\text{txt}}$.
We find that using the last-layer \texttt{[CLS]} token as feature of the TrOCR model~\cite{li2022trocr} and calculating a loss based on the $L^1$-norm of the embeddings produce robust results for this application (\autoref{fig:hamburger}).
\deleted{, as demonstrated in \autoref{fig:hamburger}.}

The placement of glyph can be manual or automatic. In the automatic case, we optimize a similarity transform per glyph to maximize silhouette coverage while avoiding overlaps and maintaining a readable text layout. We first offset each glyph by a user-specified amount to encourage padding around the text. At each optimization step, we render both \addition{a morphologically opened version of} the silhouette and glyphs into two images using white with 50\% opacity on a black background. We minimize a loss that combines (i) a coverage term $\mathcal{L}_{\mathbb{I}}$ (\refsect{sect:coverage}), (ii) an overlap cost given by $\sum \text{ReLU}(v - 0.5)$ for each pixel intensity $v \in [0, 1]$ of the rendered image and (iii) an alignment cost $\sum \| \theta \|$ that penalizes the absolute turning angles $\theta$ between consecutive glyph center-points and maintains text ordering. We note that image generation models such as DALL-E 3 \cite{BetkerDALLE} are particularly effective at generating silhouettes with a prompt, which finally results in a fully automatic calligram generation pipeline.

\begin{figure}[t]
	\centering
	\includegraphics[width=0.45\textwidth]{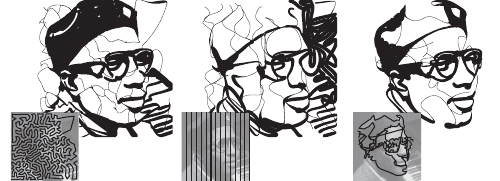}
	\caption{
      Stroke abstraction of Thelonious Monk. Left, using a single stroke and Voronoi with TSP initialzation. Middle, using multiple strokes with multiple key-points along vertical lines. Right, using facial features extracted with MediaPipe \cite{MediaPipe2019}.
} %
\label{fig:inits}
\end{figure}

\begin{figure}[t]
	\centering
	\includegraphics[width=0.45\textwidth]{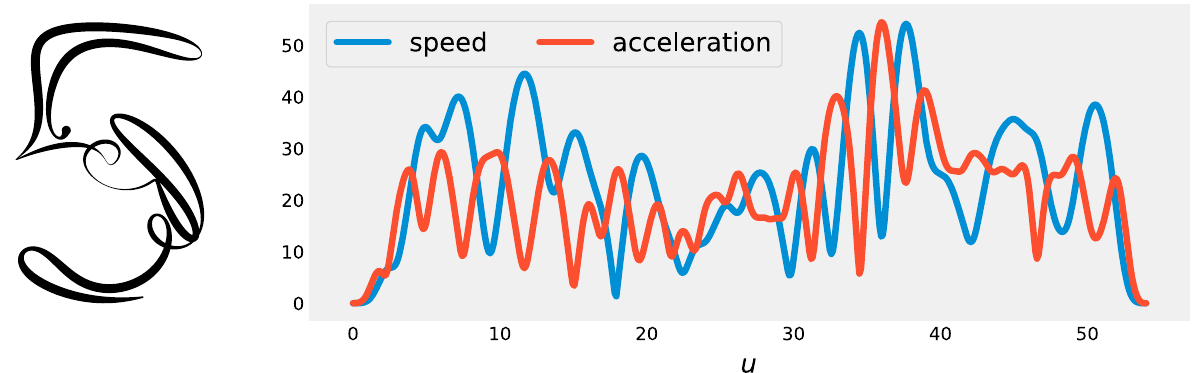}
	\caption{
Smooth speed and acceleration of a quntic spline covering a number ``5'' and optimized using the method of Section 4.1. With appropriate resampling the path kinematics can be safely tracked with a robot.
} %
\label{fig:kine}
\end{figure}
\begin{figure}[t] %
	\centering
	\includegraphics[width=0.48\textwidth]{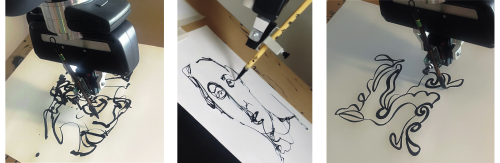}
	\caption{
      Our method produces smooth kinematics that facilitate reproduction with a robot, and the varying width can be used to control brush pressure.
      \revision{Left and center, the robot reproducing portraits. Right, the robot reproducing a stylized area fill.}
} %
\label{fig:robots}
\end{figure}

\section{Discussion}

In our example applications, we have seen how a B-spline reparame-trization can be used to \correction{enable the generation of}{generate} long and expressive strokes and curves in a DiffVG pipeline. B-splines enforce high-order continuity by design, which enables analytic smoothing losses \correction{
while also facilitating the computation of geometric costs that depend on well-behaved curve derivatives}{ that help producing more regular geometry when combined with different stylization losses}. This offers a considerable advantage compared to using only B\'ezier curves or parametrizations with lower order continuity, especially for applications like the ones demonstrated in this paper. Qualitative examples of this can be seen in examples such as \autoref{fig:smooth-comp} and  \autoref{fig:hamburger}.
\revision{For conciseness, we used a similar Voronoi-based initialization strategy in most of our examples. However, our method performs well with different initializations \revision{(\refig{fig:inits})}, which can serve as an additional design parameter to be explored by users.}

One common challenge in stroke-based abstraction pipelines is controlling the trade-off between visual fidelity and geometric simplicity. Previous methods typically address this by pre-determining the number of curves \cite{vinkerCLIPassoSemanticallyAwareObject2022} or by integrating a learned component into the optimization loop \cite{vinkerCLIPasceneSceneSketching2023}. We find that our use of smoothing, combined with optimizable stroke width, allows this trade-off to be controlled parametrically and with the number of strokes emerging from the optimization. This results in a solution that is significantly simpler than previous methods.

We investigate the utility of our representations and different loss terms in different examples of our applications. In \autoref{fig:hamburger} we perform a small qualitative ablation showing the effectiveness of the proposed legibility loss (\autoref{fig:hamburger}c) as well as the benefits of B-splines and smoothing compared to Catmull-Rom splines (\autoref{fig:hamburger}e), which only enforce $C^1$ continuity.
In \autoref{fig:smooth-comp}a we can observe that directly optimizing B\'ezier curves effectively captures features of the example style image. However, higher degrees of freedom produce results that capture finer details at the expense of a clear stroke structure. Although this additional detail may be desirable in certain applications, it is not suitable for the applications considered in our work.

Interestingly, the computational overhead of the proposed B-spline to B\'ezier matrix conversion is lower than the one for the additional optimization parameters required for an equivalent multi-B\'ezier curve. We tested performance with a simple comparison where we cover an area by optimizing the $290$ key-points of a single open cubic B-spline. We compared this to a similar setup directly optimizing the corresponding $874$ B\'ezier control points. On our hardware setup, the B\'ezier case is $4.5$ times slower. This shows that the additional cost of the proposed matrix conversion is negligible and suggests that our method is an efficient way to enforce output continuity in DiffVG settings.

\revision{
\paragraph{Robotic reproduction}
Optimizing splines with degree greater than three results in smooth acceleration profiles (\refig{fig:kine}). This enables a safe reproduction of the resulting trajectory kinematics with an articulated robot arm (\refig{fig:robots}), without requiring an intermediate  reparameterization step.
We tested this by reproducing the trajectories using a 7-axis Franka robot equipped with a brush. We first transformed the control points to a desired workspace coordinate system, treating the stroke widths as perpendicular distances to the drawing plane. We then sampled the trajectories at a resolution that produced a maximum speed and accelerations within the robot's mechanical limits. The inverse kinematics for the resulting trajectories are then computed with an iterative linear quadratic regulator (iLQR) \cite{Li04}.
}

\section{Conclusions and future work}
We have presented a framework for integrating high-order B-splines into DiffVG pipelines together with minimum-square derivative-based smoothing costs.
We have explored different applications and demonstrated how this enables the generation of long, smooth, and stylized strokes through a combination of geometric and image-space loss functions. 
While the combination of losses allows for a large variety of creative outputs, a practical challenge is the necessity to weigh different losses to achieve the desired result, which, given the iterative optimization procedure, can be slow and tedious.

Although our formulation draws on a large body of existing work on B-splines, an effective use of this tool together with DiffVG is novel, and we expect it to be a valuable tool for the community.
We used uniform B-splines because of their simplicity and effectiveness for our use cases. However, exploring non-uniform parameterizations, such as NURBs, presents an interesting direction for further research, as it may unlock additional flexibility and control for stylized outputs.

\begin{acks}
  This work was funded by the EACVA (Embodied Agents in Contemporary Visual Art) Project, led by Goldsmiths (UKRI/AHRC grant AH/X002241/1) and the University of Konstanz (grant 508324734, Deutsche Forschungsgemeinschaft/DFG). Special thanks to Guillaume Clivaz (Idiap Research Institute) for the technical support and useful discussions.
\end{acks}

\appendix
\section{B-Spline details}
A B-spline (or basis-spline) or order \(k\) is a piecewise polynomial curve of degree \(p=k-1\) defined by a linear
combination of \(n\) weights or control points \(\cp_0, \cp_1, \dots
\cp_{n-1}\) and a non-decrasing sequence of \(m=n+k\) knots (or breakpoints) \(t_0,
t_1, t_2, \dots, t_{m-1}\).

\[
  \bm{x}(u) = \sum_{i=0}^{n-1}{\cp_i B_{i,k}(u)}
\]

Each basis function \(B_{i,k}\) defines \(k\) polynomial segments spanning \(k+1\) knots \(t_i, t_{i+1}, \dots, t_{i+k}\) and is positive in the half-open domain \([t_i, t_i+k)\).
The knots between \(t_p\) and \(t_{m-k}\) (not included) are called ``internal'' or ``interior'' knots.
From \href{https://web.mit.edu/hyperbook/Patrikalakis-Maekawa-Cho/node17.html}{here}: For \(n\) control points we have \(n+k\) knots and \(n-k\) interior knots.

B-spline bases can be defined through the ``Cox-de Boor'' recursion starting from order \(1\) (degree \(0\)):
\[
B_{i,1}(u) = \begin{cases}
1 & \text{if } t_i \leq u < t_{i+1} \\
0 & \text{otherwise}
\end{cases}
\]
And with
\[
B_{i,k}(u) = \frac{u - t_i}{t_{i+k-1} - t_i} B_{i,k-1}(u) + \frac{t_{i+k} - u}{t_{i+k} - t_{i+1}} B_{i+1,k-1}(u)
\]

The number of control points \(n\), order \(k\) and number of knots \(m\) are related by \(n + k - m = 0\). For nonrepeating knot sequences, the curve will be \(C^{k-2}\) continuously differentiable.

\subsection{Derivatives}
\label{sec:orgc489b7e}
The derivative of a B-spline basis function of order \(k\) is given by
\[
\deriv{u}B_{i, k}(u)  = B^{\prime}_{i, k}(u) = \frac{k-1}{t_{i+k-1} - t_i} B_{i,k-1}(u) - \frac{k-1}{t_{i+k} - t_{i+1}} B_{i+1,k-1}(u).
\]

It is a linear combination of all the derivatives of the basis function. As a result, the derivative of a B-spline is equivalent to a B-spline of order \(k - 1\) with a new set of control points given by weighted differences of pairs of consecutive control points.

\subsection{Cardinal B-splines}
\begin{figure}[b]
	\centering
	\includegraphics[width=0.49\textwidth]{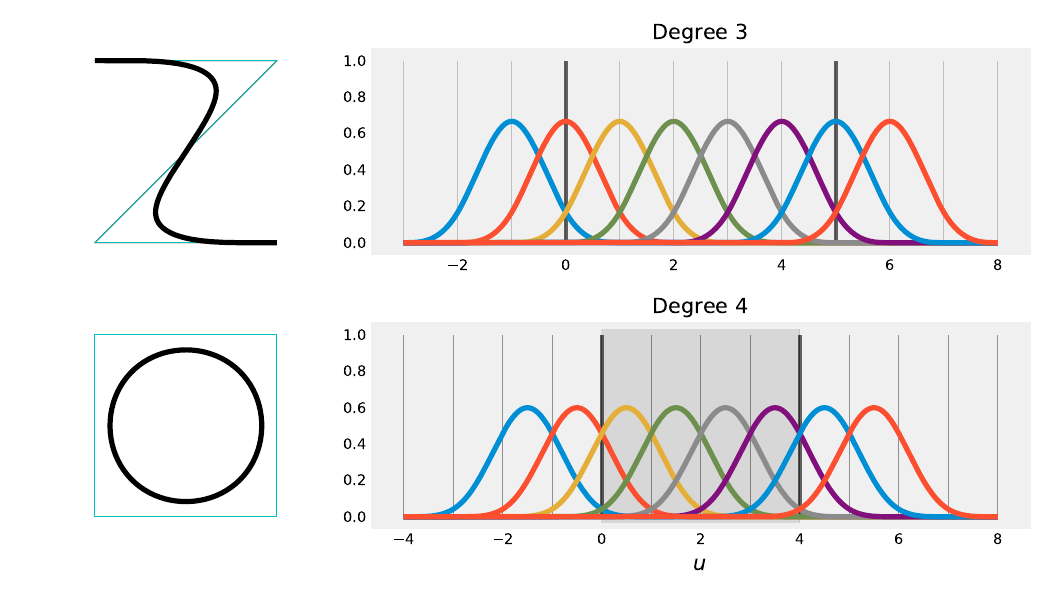}
	\caption{
    B-Splines and their bases with degrees $3$ and $4$.
} %
\label{fig:bsplines}
\end{figure}

A \emph{cardinal} B-spline (not to be confused with cardinal/Catmull-Rom splines) is a ``normalized uniform B-spline''. It has uniformly spaced knots, with \(t_{i+1} - t_i = h\) (uniform) with  \(h=1\) (normalized) so the knots are all integers (\refig{fig:bsplines}).
Uniformity and normalization simplify the computations of a B-spline as all basis functions are translated versions of the same basis function that we denote as \(N_k(u)\).
We then have

\[
\bm{x}(u) = \sum_{i=0}^{n-1}{\cp_i N_{k}(u - t_i)}
\]

and the B-spline derivatives simplify to
\[
\deriv{u}N_{k}(u)  = N^{\prime}_{k}(u) = N_{k-1}(u) - N_{k-1}(u-1)
\]

so
\[
  \dot{\bm{x}}(u) = \sum_{i=0}^{n-1}{\cp_i \left( N_{k-1}(u - t_i) - N_{k-1}(u - t_i - 1) \right)}
\]

\subsection{Smoothing term}
\label{sec:org630f290}
The smoothing term can be computed exactly and is considerably simplified for the case of cardinal B-splines \cite{Schumaker1981SplineFB}. While different approaches exist to calculate this kind of integral \cite{de1976calculating,vermeulen1992integrating}
to calculate this kind of integral, we follow
\citet{fujiokaConstructingCharacterFont2007} and \citet{fujiokaReconstructingDynamicFontbased2017} to have
\[
\mathcal{L}_{\mathrm{smooth}}^{d} = \int_{-\infty}^\infty D(u) \, \mathrm{d}u - \int_{-\infty}^{t_{k-1}} D(u) \, \mathrm{d}u - \int_{t_n}^\infty D(u) \, \mathrm{d}u
\]
with \( D(u) = \|\bm{x}^{(d)}(u)\|^2 \)

This can be computed explicitly by constructing a Gramian \(\bm{G}\) with:

\[
G_{i,j} = \begin{cases}
\int_0^k N^{(d)}_{i,j} \df{u} -
\int_0^{p-i} N^{(d)}_{i,j} \df{u}   & \text{if } i < p \text{ and } j < p\\
\int_0^k N^{(d)}_{i,j} \df{u} -
\int_0^{p-i} N^{(d)}_{n+p-i,n+p-j} \df{u}   & \text{if } i \ge n \text{ and } j \ge n\\
\int_0^k N^{(d)}_{i,j} \df{u} & \text{otherwise}
\end{cases}
\]
and
\[
N^{(d)}_{i,j} = N^{(d)}_{k}(u) N^{(d)}_{k}(u-j+i)
\].

Then each \(G_{i,j}\) can be computed exactly using quadrature
\cite{vermeulen1992integrating}.

If we let \(\bm{c}\in\reals^{nD}\) be a vector that concatenates \(n\) control points, each of dimensions \(D\) we have
\[
\mathcal{L}_{\text{smooth}}^{d} = \bm{c}^\trsp \bar{\bm{G}} \bm{c}, \quad \bar{\bm{G}}=\bm{G}\otimes\bm{I}_D
\]
where \(\otimes\) is the Kroenecker product and \(\bm{I}_D\) is the identity matrix of dimensions \(D\).

\paragraph{P-splines.} A similar procedure can be efficiently approximated with discretization of the derivative cost, using penalized-splines (P-splines) as described by \citet{eilersFlexibleSmoothingBsplines1996}.
To do this, we can simply use \(\bm{G}=\bm{D}_{(d)}^\trsp\bm{D}_{(d)}\) with \(\bm{D}_{(d)}\) a matrix representing the finite difference operator of order \(d\). The advantage of this method is the simplicity of implementation and the possibility of achieving similar smoothing results. We can arbitrarily combine the degree of discrete differences with the degree of the curve. We expose both methods for completeness and to enable applications where the integral cost may be necessary (e.g., planning and robotics).

\subsection{Conversion to B\'ezier}
With the method of \citet{romaniConversionMatrixUniform2004a}, converting the $p+1$ control points of a quintic B-spline of degree $p$  to single B\'ezier segment of the same degree, can be done with a $(p + 1) \times (p + 1)$ matrix that we denote as $\bm{S}^{p}$. To convert all the control points of a B-spline we stack multiple shifted and overlapping copies of $\bm{S}^{p}$ into a larger matrix $\bm{S}$, by shifting each copy by $p$ rows and $1$ column. For a quintic spline this can be visualized as:
\begin{equation*}
  \begin{NiceMatrix}
\mydot & \mydot & \mydot &   \mydot & \mydot & \mydot &    &  & \\
\mydot & \mydot & \mydot &   \mydot & \mydot & \mydot &    &  & \\
\mydot & \mydot & \mydot &   \mydot & \mydot & \mydot &    &  & \\
\mydot & \mydot & \mydot &   \mydot & \mydot & \mydot &    &  & \\
\mydot & \mydot & \mydot &   \mydot & \mydot & \mydot &    &  & \\
\mydot & \mydot & \mydot & \mydot & \mydot & \mydot & \mydot & & \\
      & \mydot & \mydot & \mydot & \mydot & \mydot & \mydot & & \\
      & \mydot & \mydot & \mydot & \mydot & \mydot & \mydot &\\
      & \mydot & \mydot & \mydot & \mydot & \mydot & \mydot &\\
      & \mydot & \mydot & \mydot & \mydot & \mydot & \mydot &\\
      & \mydot & \mydot & \mydot & \mydot & \mydot & \mydot &\\
\CodeAfter
\tikz \draw[gray, thick] (1-1.north west) rectangle (6-6.south east);
\tikz \draw[gray, thick] (6-2.north west) rectangle (11-7.south east);
\end{NiceMatrix}
\end{equation*}
The blocks for a quintic spline are given by:
\[
\bm{S}^{5} = \frac{1}{120}
\begin{bmatrix}
1 & 26 & 66 & 26 & 1 & 0 \\
0 & 16 & 66 & 36 & 2 & 0 \\
0 & 8  & 60 & 48 & 4 & 0 \\
0 & 4  & 48 & 60 & 8 & 0 \\
0 & 2  & 36 & 66 & 16 & 0 \\
0 & 1  & 26 & 66 & 26 & 1 \\
\end{bmatrix}.
\]
The block matrix $\bar{\bm{S}}$ used to compute the B\'ezier control points from the flattend B-spline control points $\bm{c}$ is given by the Kroenecker product $\bm{S} \otimes \bm{I}_{D}$.

\subsection{Degree reduction}
With the method of \citet{sunwooMatrixRepresentationMultidegree2005}, reducing a B\'ezier curve of degree $p$ to one of degree $q$ can be done with a $(q + 1)  \times (p + 1)$ matrix that we denote as $\bm{R}^{p,q}$. To reduce the degree of all the control points of a B\'ezier chain we stack multiple shifted and overlapping copies of $\bm{R}^{p,q}$ into a larger matrix $\bm{R}$ by shifting each copy by $p$ rows and $q$ columns. For a reduction from quintic to cubic this can be visualized as:

\setcounter{MaxMatrixCols}{30}

\begin{equation*}
\begin{NiceMatrix}
       \mydot & \mydot & \mydot &   \mydot & \mydot & \mydot &        &        &        &        & \\
       \mydot & \mydot & \mydot &   \mydot & \mydot & \mydot &        &        &        &        & \\
       \mydot & \mydot & \mydot &   \mydot & \mydot & \mydot &        &        &        &        & \\
       \mydot & \mydot & \mydot &   \mydot & \mydot & \mydot & \mydot & \mydot & \mydot & \mydot & \mydot \\
              &        &        &          &        & \mydot & \mydot & \mydot & \mydot & \mydot & \mydot \\
              &        &        &          &        & \mydot & \mydot & \mydot & \mydot & \mydot & \mydot \\
              &        &        &          &        & \mydot & \mydot & \mydot & \mydot & \mydot & \mydot \\
\CodeAfter
\tikz \draw[gray, thick] (1-1.north west) rectangle (4-6.south east);
\tikz \draw[gray, thick] (4-6.north west) rectangle (7-11.south east);
\end{NiceMatrix}
\end{equation*}

The blocks of the quintic to cubic reduction matrix are given by
\[
\bm{R}^{5,3} = \begin{bmatrix}
1 & 0 & 0 & 0 & 0 & 0 \\
-\frac{2}{3} & \frac{5}{3} & 0 & 0 & 0 & 0 \\
0 & 0 & 0 & 0 & \frac{5}{3} & -\frac{2}{3} \\
0 & 0 & 0 & 0 & 0 & 1
\end{bmatrix}
\]

The block matrix $\bar{\bm{R}}$ used to compute the reduced B\'ezier control points from is given by the Kroenecker product $\bm{R} \otimes \bm{I}_{D}$.

\bibliographystyle{ACM-Reference-Format}
\bibliography{calligraph-paper}

\end{document}